\documentclass[12pt,thmsa]{article}
\usepackage{amsmath}
\usepackage{amsfonts}
\usepackage{amssymb}
\usepackage{graphicx}

\input{tcilatex}
\begin{document}

\begin{center}
{\Large Berry phase in superconducting charge qubits interacting with a
cavity field \bigskip }

M. Abdel-Aty

{\small Mathematics Department, Faculty of Science, Sohag
University, 82524 Sohag, Egypt }
\end{center}

{\small We propose a method for analyzing Berry phase for a multi-qubit
system of superconducting charge qubits interacting with a microwave field.
By suitably choosing the system parameters and precisely controlling the
dynamics, novel connection found between the Berry phase and entanglement
creations.}

{\small PACS numbers: 74.70.-b, 03.65.Ta, 03.65.Yz, 03.67.-a, 42.50.-p}

\section{Introduction}

Berry's phase \cite{ber84} can be thought of as an adiabatic quantum
holonomy restricted to a one-dimensional energy eigenspace \cite{sjo08}. The
Berry phase has very interesting applications, such as the implementation of
quantum computation by geometric means \cite{jon99,eke00,fal00}. Recently,
it has been recognized that large-scale quantum computers are hard to
construct because quantum systems easily lose their coherence through
interaction with the environment \cite{sjo08}. Researchers have tried to
avoid this problem by using geometric phase shifts in the design of quantum
gates to perform information processing. The robustness of Berry's geometric
phase for spin-1/2 particles in a cyclically varying magnetic field has been
tested experimentally \cite{fil09}. Interesting Berry phase results of a
composite system have also been presented in \cite{yi04} and a new formalism
of the geometric phase for mixed states in the experimental context of
quantum interferometry has been discussed in Ref. \cite{sjo00}. In another
set of experiments, Du et al. \cite{du03} performed an NMR experiment to
measure the geometric phase for mixed quantum states, and Leek et al. \cite%
{lee08} analyzed experimentally the Berry phase for a superconducting qubit
affected by parameter fluctuations.

Another important problem associated with quantum computation and
information is the problem of engineering entanglement in multi-particle
systems. This has attracted a great deal of recent interest \cite%
{fac09,fac08,guh08,oh08}. Here questions of how the couplings among the
subsystems changes the Berry phase of the composite system become of great
importance. In this regard, the relation between entanglement and the Berry
phase has been discussed in solid state systems \cite{ryu06} and in
icosahedral Jahn-Teller systems \cite{lij05}. Most of the earlier works on
the geometrical phase focus on pure quantum states \cite%
{pat98,gar98,mos99,sjo97,jai98}. These types of systems, however, are very
unrealistic and almost never occur in practice. In some applications,
however, in particular geometric fault tolerant quantum computation \cite%
{jon99}, mixed state cases are important. From a mathematical point of view,
Uhlmann was the first to address the issue of mixed state holonomy \cite%
{uhl86,uhl91}. The difficulty with mixed states is their reduced coherence,
which makes any notion of a phase more difficult to define and measure.
Another challenge for the future is to extend this work to more general
forms of mixed state multi-qubit systems.

The aim of the present work is to develop a general treatment of the
multi-qubit problem and develop realizable procedures to various base
spaces. This is dome treating the non-degenerate time evolution operator and
satisfying parallel transport evolution condition and by defining the Berry
phase in a new, rather simple, way. In particular, one of the most important
problems under consideration is how to define the Berry phase of multi-qubit
system and easy-monitored entangled state with existing experimental
techniques \cite{vid00}. We will show that setting up an arrangement in
which an entanglement can be measured via the Berry phase for
superconducting qubits is an interesting task both from the experimental and
theoretical viewpoints.

\section{Mixed state Berry phase}

The Berry phase has been extensively studied \cite{yi04,geo89,sve94,muk93}
and generalized in various ways. For example, based on the fact that any
mixed state can be represented as a pure one if one allows ancillas and
optimizes over many purification, the mixed state geometric phase has been
defined \cite{sjo00}, for open systems \cite{car03}, and with a quantized
field driving \cite{fue02}.

To study the geometric properties of a quantum system \cite{car04}, we
evaluate the Berry phase of the system by introducing a phase shift operator

\begin{equation}
R(t)=\exp [-i\phi (t)\psi ^{\dag }\psi ],
\end{equation}%
where $\phi (t)$ is changed from $0$ to $2\pi $ adiabatically. We denote by $%
\psi ^{\dagger }(\psi )$ the creation (annihilation) operator of the cavity
mode. Then the time independent eigenequation of the system $H|\chi
_{j}\rangle =\Im _{j}|\chi _{j}\rangle ,$ is changed into $H^{\prime }|\chi
_{j}^{\prime }\rangle =\Im _{j}^{\prime }|\chi _{j}^{\prime }\rangle $, with
\begin{equation}
H^{\prime }=R(t)HR^{\dag }(t)-iR(t)\frac{dR^{\dag }(t)}{dt},
\end{equation}%
and $|\chi _{j}^{\prime }\rangle =R(t)|\chi _{j}\rangle $. Hence, the Berry
phase generated after the system undergoing an adiabatic and cyclic
evolution may be calculated as follows
\begin{equation}
\gamma _{j}(\tau )=i\int\limits_{0}^{\tau }\langle \chi _{j}|R^{\dag }(t)%
\frac{dR(t)}{dt}|\chi _{j}\rangle dt.  \label{be1}
\end{equation}%
Whenever a pure quantum state undergoes a parallel transport along a closed
path, it gathers information on the geometric structure of the Hilbert space
in which it lies. On the other hand, a state of subsystem is no longer a
pure one and to study the Berry phase of the subsystem (a non-degenerate
density matrix), we have to adopt the definition of geometric phase for a
mixed state \cite{sjo00}. Our goal in the following is to establish a
connection between the Berry phase acquired by a multi-qubit system and the
Berry phases of the subsystem.

The geometric phase corresponding to the non-unitary evolution can be
defined according to the geometric phase of the whole system that evolves
unitarily. For example, let us consider the following mixed state%
\begin{equation}
\rho (t)=\sum\limits_{k=1}^{L}\xi _{k}U(t)\rho _{k}(0)U^{\dag }(t),
\label{den1}
\end{equation}%
where $\rho _{k}(0)$ is an orthonormal set of pure states and $\xi _{k}$ is
the probability, where, $\rho (0)=\sum_{k=1}^{L}\xi _{k}\rho _{k}(0)$ and $L$
is the total number of the involved pure states. For non-degenerate $\xi _{k}
$ the time evolution operator is given by $U(t)=U_{0}(t)\sum_{k}e^{-i\alpha
_{k}}$ $\rho _{k}(0),$ where $\alpha _{k}$ are arbitrary real parameters and
$U_{0}(t)$ is one of the equivalent operators to $U(t)$. When a mixed state
given by the density matrix $\rho (t)$ evolves under a unitary operator $%
U(t),$ the parallel transport evolution condition is
\begin{equation}
Tr[\rho _{k}(t)\overset{.}{U}(t)U^{\dag }(t)]=0,\ (k=1,2,...,L).
\label{den2}
\end{equation}%
Based on geometric phase definition for a mixed state \cite{sjo00} and using
Eqs. (\ref{den1}) and (\ref{den2}), for an adiabatic cyclic evolution, a new
definition of the Berry phase for a mixed states can be written as
\begin{eqnarray}
\gamma _{B}(\tau ) &=&\arg \left\{ \sum\limits_{k=1}^{L}\text{Tr}\left[ \xi
_{k}\rho _{k}(0)U_{1}(\tau )\right. \right.   \notag \\
&&\left. \left. \times \exp \left( -\int\limits_{0}^{\tau }\text{Tr}[\rho
_{k}(0)U_{0}^{\dag }(t)\overset{.}{U}_{0}(t)]dt\right) \right] \right\} .
\label{Be1}
\end{eqnarray}%
The Berry phase Eq. (\ref{Be1}) for a mixed state is just an average of the
individual Berry phases. The Berry phase factor for a mixed state defined by
Eq. (\ref{Be1}) is a weighted sum of the one-particle Berry phase factors.
In the pure state case Eq. (\ref{Be1}) is consistent with the result of Eq. (%
\ref{be1}).

\subsection{Multi-qubit system}

Over the last decade, superconducting qubits have gained substantial
interest as an attractive option for quantum information processing \cite%
{you05,mak01,pas05} and Josephson qubits are among the most promising
devices to implement solid state quantum computation \cite{rom05,ave00}. A
novel method for the controlled coupling of two Josephson charge qubits by
means of a variable electrostatic transformer has been proposed in Ref. \cite%
{ave03} and the behavior of charge oscillations in superconducting Cooper
pair boxes weakly interacting with an environment has been discussed \cite%
{ben08}. Also, the quantum dynamics of a Cooper-pair box with a
superconducting loop in the presence of a nonclassical microwave field have
been investigated in Ref. \cite{you03}.

\begin{figure}[tbph]
\begin{center}
\includegraphics[width=16cm,height=7cm]{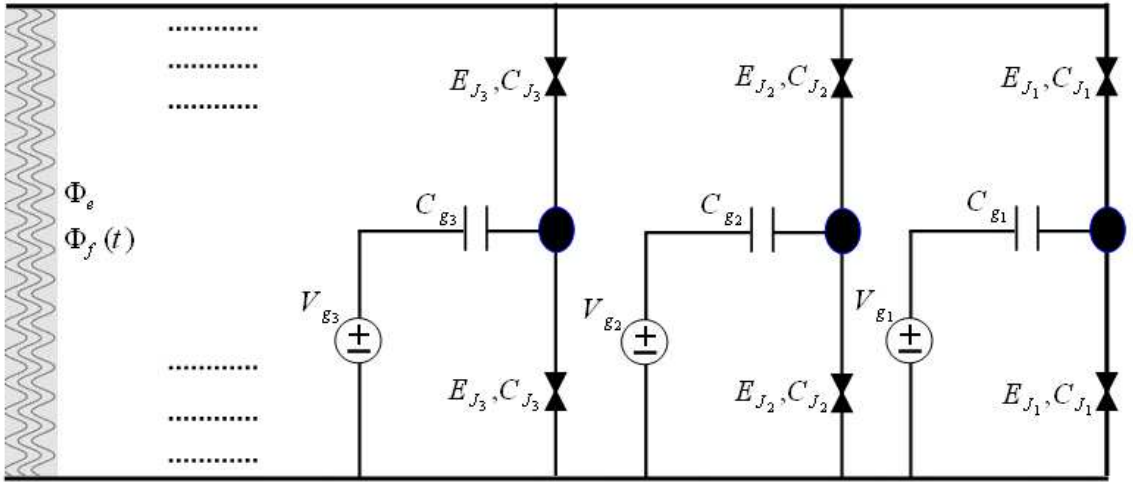}
\end{center}
\caption{Schematic picture of the multi-Cooper-pair-box system. The filled
circle denotes a superconducting island or the Cooper-pair box. It is biased
by a voltage $V_{g_{i}}$ through the gate capacitance $C_{g_{i}}$ and
coupled to the bulk superconductors by two identical small Josephson
junctions. The two Josephson junctions have capacitance $C_{J_{i}}$ and
Josephson energy $E_{Ji}$. The total flux is the summation of the static
magnetic flux $\Phi _{e}$, and microwave-field-induced flux $\Phi _{f}(t)$
applied via the SQUID loop.}
\end{figure}

As an explicit example, we consider a suitable multi-qubit system and
envisage a process in which the qubits interact with the microwave cavity
such that there is no overlap of their interactions. Let us start with a
short description of a superconducting box formed from a low-capacitance
Josephson junction of capacitance $C_{J_{i}}$ and Josephson energy $%
E_{J_{i}}.$ The Josephson junction is biased by a voltage source $V_{g_{i}}$
through a gate capacitance $C_{g_{i}}$ which is externally controlled and
used to induce offset charges on the island. The schematic picture of this
multi-qubit structure is shown in figure 1. The total Hamiltonian of the
single-Cooper pair box system can be written as \cite{mig01}
\begin{eqnarray}
\hat{H} &=&\hbar \omega _{k}(\widehat{\psi }^{\dagger }\widehat{\psi }+\frac{%
1}{2})  \notag \\
&&+\frac{(Q-C_{g_{i}}V_{g_{i}})^{2}e}{C_{g_{i}}+2C_{J_{i}}}-2E_{J0}\cos \phi
\cos \left( \pi \frac{\Phi }{\Phi _{0}}\right) ,  \label{5}
\end{eqnarray}%
where $\widehat{\psi }^{\dagger }(\widehat{\psi })$ is the creation
(annihilation) operator of the cavity mode. In this structure, the
superconducting island with Cooper-pair charge $Q=2Ne$ is coupled to a
segment of a superconducting ring via two Josephson junctions, where $e$ is
the electron charge and $N$ \ is the number of Cooper-pairs. We denote by $%
\phi =0.5(\phi _{1}+\phi _{2})$ the phase difference across the junction.
The gauge-invariant phase drops $\phi _{1}$ and $\phi _{2}$ across the
junctions are related to the total flux $\Phi $ through the SQID loop by the
constraint $\phi _{2}-\phi _{1}=2\pi \Phi /\Phi _{0},$ where $\Phi _{0}=h/2e$
is the quantum flux. We assume that the structure of the Cooper pair box is
characterized by two energy scales, i.e., the charging energy $E_{c}$ and
the coupling energy $E_{J_{0}}$ of the Josephson junction, and we consider
that the charging energy with scale $Ec=e^{2}/2(C_{g}+C_{J})$ is chosen to
dominate over the Josephson coupling energy $E_{J}$ and weak quantized
radiation field, so that only the two low-energy charge states $N=0$ ($%
\left\vert 0\right\rangle $) and $N=1$($\left\vert 1\right\rangle $) are
relevant \cite{mak99} while all other charge states, having a much higher
energy, can be ignored. When a nonclassical microwave field is applied, the
total flux is a quantum variable $\Phi =\Phi _{e}+\Phi _{f}(t),$ where $\Phi
_{f}$ is the microwave-field-induced flux. If we consider a planar cavity
containing the superconducting-quantum-interference-device loop of the
charge qubit perpendicular to the cavity mirrors, the vector potential of
the nonclassical microwave field can be written as $A(r)=|u_{k}(r)|(\widehat{%
\psi }^{\dagger }+\widehat{\psi })\widehat{A},$ where a single-qubit
structure is embedded in the microwave cavity with only a single photon mode
$\lambda .$ Thus, the flux $\Phi _{f}$ can be written as $\Phi _{f}=|\Phi
_{k}|(\widehat{\psi }^{\dagger }+\widehat{\psi }),$ where $\Phi _{k}=\oint
u_{k}.dl.$ We shift the gate voltage $V_{g}$ (and/or vary $\Phi _{e}$) to
bring the single-qubit system into resonance with $k$ photons: $E\approx
k\hbar \omega _{k}$, $k=1,2,3,....$ Note that the charge states are not the
eigenstates of the Hamiltonian (\ref{ham}), so that the Hamiltonian can be
diagonalized yielding the following two charge states $|e\rangle =\cos \eta
\left\vert 1\right\rangle -\sin \eta \left\vert 0\right\rangle $ and $%
|g\rangle =\sin \eta \left\vert 1\right\rangle +\cos \eta \left\vert
0\right\rangle $ with $\eta =\frac{1}{2}\tan ^{-1}(E_{J}/2\varepsilon ),$
where $\varepsilon =2E_{ci}(C_{gi}V_{gi}e^{-1}-(2n+1)).$ Employing these
eigenstates to represent the qubits, expanding the functions $\cos (\pi \Phi
/\Phi _{0})$ and using the rotating wave approximation, one can derive the
total Hamiltonian of the system as \cite{you05,mak99}
\begin{eqnarray}
\hat{H} &=&\hbar \omega _{k}(\widehat{n}+\frac{1}{2})  \notag \\
&&+\sum\limits_{j=1}^{m}\left\{ \frac{1}{2}E-E_{J0}^{(j)}\sin \left( 2\eta
\right) \cos \left( \frac{\pi \Phi _{e}}{\Phi _{0}}\right) f(\widehat{n}%
)\right\} \sigma _{z}^{(j)}  \notag \\
&&+\sum\limits_{j=1}^{m}\left\{ E_{J0}^{(j)}\cos \left( 2\eta \right) \psi
^{k}g_{_{(k)}}(\widehat{n})\sigma _{+}^{(j)}+H.c.\right\} .  \label{ham}
\end{eqnarray}%
Here we denote by $\sigma _{\pm }^{(j)}$ and $\sigma _{z}^{(j)}$ the Pauli
matrices in the pseudo-spin basis of the $j^{th}$ qubit and $g_{k}(\widehat{n%
}),$ ($\widehat{n}=\widehat{\psi }^{\dagger }\widehat{\psi }$) represents
the $k-$photon-mediated coupling between the charge qubit and the microwave
field.

The influence of entanglement on the noncyclic two-particle geometric phase
has been studied for an entangled spin pair in Refs. \cite{sjo00,sjo05,ge05}%
. It was shown that prior entanglement shared between the two spins
can change the Berry phase in such an entangled pair. Also, an
experimental technique for preparing arbitrarily entangled
polarization states has been developed \cite{whi99}. The treatment
here extends these investigations \textrm{to explore, in a
controlled setting, the fundamental features of the relation between
Berry-phase and entanglement }in multi-qubit system. In particular,
we consider a process in which multiple superconducting charge
qubits are placed in parallel and in addition to a static magnetic
flux, a microwave field is applied through the SQUID loop. Also,
each qubit interacts with the microwave field such that there is no
overlap of their interaction with the field or with each other
\cite{dat04,gho06}. \textrm{In a similar model \cite{you02},
multiple charge qubits are placed in parallel and coupled via a
common inductance. }

\subsection{Numerical results}

\begin{figure}[tbph]
\begin{center}
\includegraphics[width=8cm]{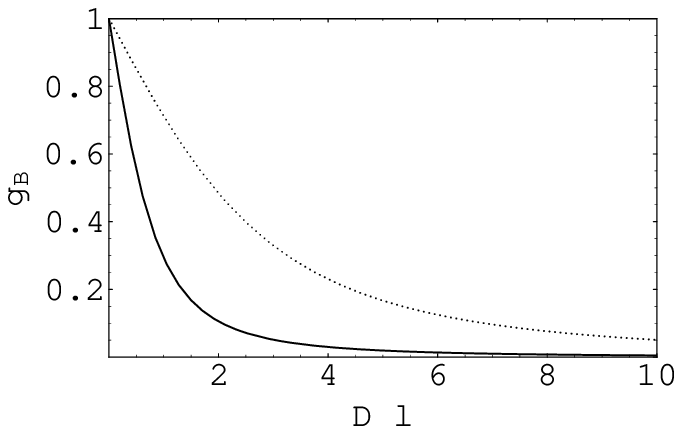} \includegraphics[width=8cm]{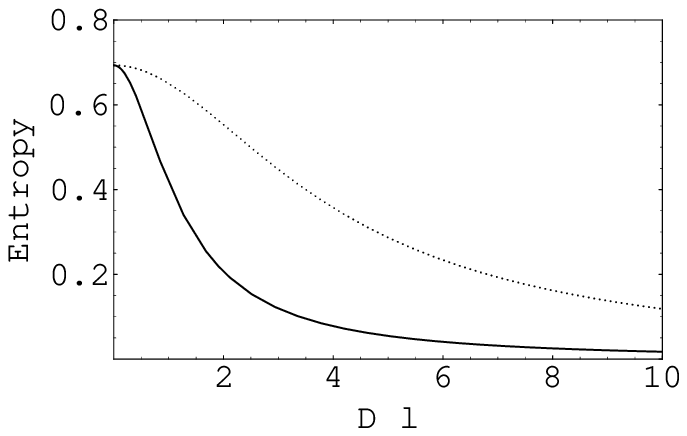}
\end{center}
\caption{ Berry phase $\protect\gamma _{B}(t)$ and entropy corresponding to
the first instantaneous eigenstate versus the dimensionless detuning
parameter $\Delta /\protect\lambda $ $(\protect\lambda =\protect\sqrt{e^{2}%
\protect\omega /(\hbar C_{F}}),$ for different number of n. Here the curves
are for $n=0$ (solid curve) and $n=10$ (dotted curve). The plot was
normalized and presented in units of \ $\protect\pi $\ for the Berry phase.}
\label{berry1}
\end{figure}
In order to understand a number of different cases, we show, in what
follows, plots of the Berry phase for different values of the involved
parameters. As might be expected, the behavior of the present system changes
depending on the number of Cooper pair boxes (qubits) involved, the
variation of the system parameters, i.e., the Josephson energy $E_{J_{i}},$
gate capacitance $C_{g_{i}}$ and initial state of the field. The Berry's
phase defined in equation (\ref{Be1}) is calculated using quantum states
evolving in time under the action of Hamiltonian $\hat{H}$. \ If the
evolution of only one qubit is considered, it is straightforward to write
down the formal solution of the problem as $\hat{H}|s,g\rangle =\Omega
_{0}|s,g\rangle ,\quad 0\leq s<k$ \ and $\hat{H}|\Phi _{\pm }^{(n)}\rangle
=\Omega _{\pm }^{(n)}|\Phi _{\pm }^{(n)}\rangle ,$ where $\Omega _{0}$ and $%
\Omega _{\pm }^{(n)}$ are the eigenvalues. The entangled states $|s,g\rangle
$ and $|\Phi _{\pm }^{(n)}\rangle $ are orthonormal and complete. Here $\pm $%
, labels the entangled states which in their bare condition are the ground $%
|g\rangle $ and the excited $|e\rangle $ states, and $n$, the states of the
boson field. The unitary operator $\hat{U}(t)$ can be written as
\begin{eqnarray}
\hat{U}(t) &=&\sum\limits_{n=0}^{\infty }\sum\limits_{l=+,-}\left\{ \exp
(-it\Omega _{l}^{(n)})|\Psi _{l}^{(n)}\rangle \langle \Psi
_{l}^{(n)}|\right\}  \notag \\
&&+\sum\limits_{s=0}^{k-1}\exp (-it\Omega _{0})|s,g\rangle \langle g,s|.
\label{ut}
\end{eqnarray}%
In Fig. (\ref{berry1}), we use Eq. (\ref{ut}) and assume that the qubits
initially prepared in the excited level interact with the cavity under
consideration are in the pure number state (Fock state), i.e. $\hat{\rho}%
_{f}\left( 0\right) =|n\rangle \langle n|$. In this figure we show the
measured Berry phase $\gamma _{B}(\tau )$ and its dependence on the scaled
detuning $\Delta /\lambda ,$ ($\lambda =\sqrt{e^{2}\omega /(\hbar C_{F}})$,
for different value of $n$. The calculations are all carried out with the
single photon process and $\Phi _{e}=\Phi _{0}/2$ for all Josephson charge
qubits. Here two parameters are varied; the number $n$ and the dimensionless
detuning parameter $\Delta /\lambda $. The measured phase is in all cases
seen to be exponentially decay with the development of the detuning.
Detuning is expected to have the same influence on Berry's phase and
entanglement. This characteristic robustness of Berry phases may be
exploitable in the realization of logic gates for quantum computation. To
compare the Berry phase with entanglement, we use the von Neumann entropy, $%
S(t)=Tr[\hat{\rho}\left( t\right) \ln \hat{\rho}\left( t\right) ]$. The
quantum dynamics described by the Hamiltonian (\ref{ham}) leads to an
entanglement between the field and the qubits. We can express the entropy $%
S(t)$ in terms of the eigenvalue $\Upsilon _{i}(t)$ of the reduced density
operator \cite{vid00}, as $S(t)=-\sum_{i}\Upsilon _{i}(t)\ln \Upsilon
_{i}(t).$ For a single qubit, disentangled pure joint state leads to $S(t)$
is zero, and for maximally entangled states the entropy gives $\ln 2$.
\textrm{\ In our consideration to the behavior of the entropy as a function
of the detuning parameter }$\Delta /\lambda $\textrm{. When we take }$\Delta
/\lambda =0$\textrm{, we get almost maximum value for the entropy. As the
detuning parameter is increased, the entropy as well as the Berry phase are
decreased. Further increasing of the detuning leads to vanishing of both }$%
\gamma _{B}(\tau )$ and $S(t),$\textrm{\ which means a completely pure state
is reached (see figure 2).}

In the two qubits case, the eigenstates can be written as
\begin{equation}
|\psi _{n}^{(j)}\rangle =\sum_{i=1}^{4}a_{i}^{(j)}(n)|\Phi _{i}\rangle ,
\end{equation}%
where ($|\Phi _{i}\rangle =|e,e,n\rangle ,|e,g,n+k\rangle ,|g,e,n+k\rangle
,|g,g,n+2k\rangle ),$ which have an additional phase comes from a
geometrical feature \cite{che08}, i.e., the Berry phase is given by $
%\begin{equation}
$%
\begin{equation}
\gamma _{n}=2\pi (|a_{1}(n)|^{2}-|a_{4}(n)|^{2}).  \label{dens}
\end{equation}%
For the density matrix $\hat{\rho}(t)=|\psi _{n}^{(j)}\rangle \langle \psi
_{n}^{(j)}|,$ which represents the state of a bipartite system, concurrence
is defined as
\begin{equation}
C(\hat{\rho})=\max [0,\lambda _{1}-\lambda _{2}-\lambda _{3}-\lambda _{4}],
\label{coc}
\end{equation}%
where the $\lambda _{i}$ are the non-negative eigenvalues, in decreasing
order ($\lambda _{1}\geq \lambda _{2}\geq \lambda _{3}\geq \lambda _{4}$),
of the Hermitian matrix ${\Upsilon }\equiv \sqrt{\sqrt{\hat{\rho}}\widetilde{%
\rho }\sqrt{\hat{\rho}}}$ and $\widetilde{\rho }=\left( {\sigma }_{y}\otimes
{\sigma }_{y}\right) \hat{\rho}^{\ast }\left( {\sigma }_{y}\otimes {\sigma }%
_{y}\right) $. Here, $\hat{\rho}^{\ast }$ represents the complex conjugate
of the density matrix $\hat{\rho}$ when it is expressed in a fixed basis and
${\sigma }_{y}$ represents the Pauli matrix in the same basis. The function $%
C(\hat{\rho})$ ranges from $0$ for a separable state to $1$ for a maximum
entanglement. Using Eq. (\ref{dens}), the corresponding concurrences are
given by
\begin{equation}
C_{n}=2\max (0,|a_{1}(n)a_{3}(n)-a_{2}(n)|^{2}).
\end{equation}%
The concurrences $C_{n}$ range from $0$ (an unentangled product state) to $1$
(a maximally entangled state).

\begin{figure}[tbph]
\begin{center}
\includegraphics[width=9cm]{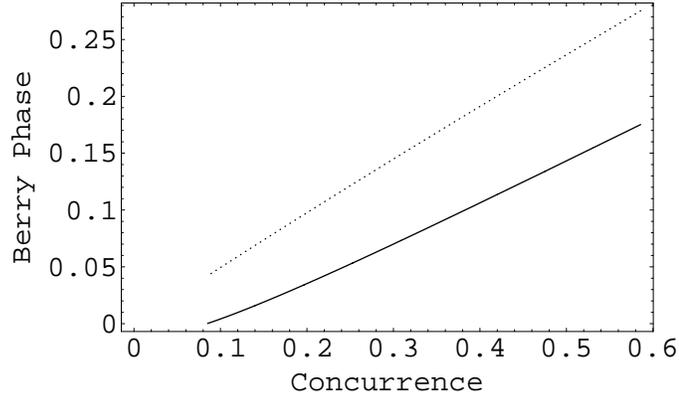}
\end{center}
\caption{ Relations between Berry phase $\protect\gamma _{B}(t)$ and
concurrences $C_{n}$ of the first eigenstate varying $n$ from zero for 10
and for different values of the detuning parameter $\Delta /\protect\lambda %
, $ where, the solid curve indicates the off-resonant case $\Delta /\protect%
\lambda =0.3$, while the dotted curve shows the resonant case $\Delta /%
\protect\lambda =0.$ }
\label{rel}
\end{figure}
Results are reported in Fig. (\ref{rel}) for the relationship between the
Berry phase and concurrence. From this figure, we confirm numerically that
there is a linear relationship between Berry phase and concurrence which
means that Berry phase is maximum when an eigenstate become a maximally
entangled state. In a special case, if we consider an entangled Bell state
given by
\begin{equation}
|\psi _{Bell}\rangle =\frac{1}{\sqrt{2}}\left( |e,e\rangle \pm |g,g\rangle
\right) ,
\end{equation}
the Berry phase factor gives exactly the measure of formation of
entanglement which is usually given by the concurrence \cite{bas06}.
Finally, we would like to make a few remarks on the relation between Berry
phase and entanglement. Berry's phase has a classical analogue: Hannay's
angle \cite{han85} is a phase effect in a classical periodic system that
depends on adiabatically changing parameters. When the Hannay angle of a
system depends on its action, the corresponding quantum system acquires a
Berry phase during the same cyclic evolution \cite{ber85}. Also,
Aharonov-Bohm effects have no classical analogue, but we may treat it as an
example of Berry's phase. More generally, however, the Aharonov-Bohm and
Berry phases can combine in a topological phase \cite{aha94}.

\section{Conclusions}

Based on the above general analysis, the essence of the Berry phase may be
summarized as follows: The present study obtains explicit results for the
Berry phase for a multi-qubit system (superconducting charge qubits). We
considered the charge-qubit with a SQUID loop and used the microwave field
to change the flux through the loop. \textrm{We have illustrated the
relation between the Berry phase and entanglement by examining different }%
examples in which the Berry phase behaves similar to the entanglement. We
have shown an interesting phenomenon of delayed Berry phase due to larger
detuning that initially entangled junction and field become sparable after a
finite time. Thus it provides an excellent basis for a further analysis of
the interplay between Berry phase and entanglement. Finally, one may say
that, the superconducting qubits model appears quite promising both as a
nice theoretical tool, with unusual access to exact analytical developments,
and in view of physical implementations in quantum information processor,
including the realization of complex single-qubit manipulation schemes and
the generation of entangled states. Also, this would help elucidate the
relation between multipartite entanglement and the Berry phase. \bigskip

%\begin{acknowledgments}

\textbf{Acknowledgments:} I acknowledge fruitful discussions with
Dimitris Tsomokos, A Bouchene and Arthur McGurn.
%This work was
%partially supported by Universit\'{e} Paul Sabatier, France.
%I
%gratefully acknowledges a visit to ICTP, Italy during which this
%work started.
%\end{acknowledgments}

\ \ \

\end{document}